\documentclass[12pt]{article}
\usepackage{epsfig}

\def\beq{\begin{equation}}
\def\eeq#1{\label{#1}\end{equation}}
\def\eeqn{\end{equation}}

\def\beqa{\begin{eqnarray}}
\def\eeqa#1{\label{#1}\end{eqnarray}}
\def\eeqan{\end{eqnarray}}

\let\bar=\overbar

\def\etal{{\it et al.}}
\def\ie{{\it i.e.}}

\def\L{{\cal L}}

\def\Dslash{\not{\hbox{\kern-4pt $D$}}}
\def\dslash{\not{\hbox{\kern-2pt $\del$}}}

\def\ee{e^+e^-}

\def\msb{{\bar{\ssstyle M \kern -1pt S}}}

\def\Title#1{\begin{center} {\Large {\bf #1} } \end{center}}

\begin{document}

\Title{Charmonium Physics with PANDA at FAIR}

\begin{center}{\large \bf Contribution to the proceedings of HQL06,\\
Munich, October 16th-20th 2006}\end{center}

\bigskip\bigskip


\begin{raggedright}  

{\it James Ritman\index{Ritman, J.} for the PANDA collaboration\\
Institute for Nuclear Physics (IKP)\\
Forschungszentrum J\"ulich\\
D-52425  J\"ulich, Germany}
\bigskip\bigskip
\end{raggedright}

\section{Introduction}
The science goals underlying the future international Facility for Antiproton and Ion Research - FAIR - \cite{FAIR} that is being realized in Darmstadt span a broad range of research activities on the structure of matter. One component of this facility is directed towards studies of hadronic matter at the sub-nuclear level with beams of antiprotons. These studies focus on two key aspects: confinement of quarks and the generation of the hadron masses. These goals will be pursued by performing precision measurements of charged and neutral decay products from antiproton-proton annihilation in the charmonium mass region.

The PANDA experiment, located at an internal target position of the High Energy Storage Ring for anti-protons HESR, is one of the large installations at the future FAIR facility~\cite{PANDA}. It is being planned by a multi-national collaboration, currently consisting of about 350 physicists from 50 institutions in 15 countries. The PANDA detector is designed as a multi-purpose setup.  The cornerstones of the PANDA physics program are:
\begin{itemize}
\item Study of narrow charmonium states with unprecedented precision
\item Search for gluonic excitations such as hybrids and glueballs in the charmonium mass region
\item Investigate the properties of mesons with hidden and open charm in the nuclear medium
\item Spectroscopy of double strange hypernuclei.
\end{itemize}
In addition to these topics a number of additional physics opportunties will open up as the facility achieves or exceeds the design goals for luminosity and resolution. These include electromagentic final states (e.g. wide angle Compton Scattering, EM Formfactors, etc.), D-meson spectroscopy, excited strange and charm baryon spectroscopy as well as CP violation in the D and/or $\Lambda$ sectors.

In this report I will concentrate on the charmonium issues relevant to PANDA.

\section{Charmonium physics with PANDA}

 The discovery of the $J/\Psi$ in 1974 and subsequently other charmonium systems greatly stimulated the understanding of the strong interaction in terms of QCD. The low density of states and the small widths below the open charm threshold reduce mixing among them thereby offering unique advantages for understanding quarkonium.

\begin{figure}[htb]
\begin{center}
\epsfig{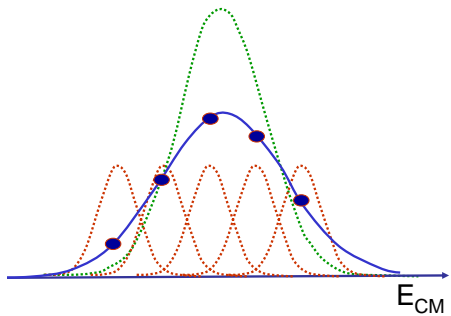}
\caption{Schematic overview of the resonance scan technique.}
\label{fig:scan}
\end{center}
\end{figure}

Extensive studies of the $\Psi$ states have been performed at $\ee$ machines where they can be formed directly. In contrast, formation reactions of the type $\bar p p \to X$ can excite charmonium states of all quantum numbers. As a result, the precision of mass and width measurements is determined by the precision of the phasespace cooled beam momentum distribution and not the (significantly poorer) detector resolution. The concept of the resonance scan is illustrated in Figure~\ref{fig:scan}.
In this figure the mass distribution that we would like to measure for the particle of interest is indicated by the green solid curve. The nominal beam momentum is adjusted to several discrete values. Due to the finite momentum distribution of the beam, each nominal setting of the beam momentum excites a distribution of $\sqrt{s}$ as indicated by the red dashed curves. The measured rate of a given final state is the convolution of these two distributions, as indicated by the filled points along the blue line. The power of this method is clearly seen by comparing measurements of the total decay width of the $\chi_{c1}$. Measurements from the Crystal Ball were only able to achieve a precision $< 3.8 MeV$~\cite{CBchic1} and newer measuremnts from BES achieved precisions of $\Gamma=1.39^{+0.40}_{-0.38} \, ^{+0.26}_{-0.77} MeV$~\cite{BESchic1}. In contrast, E835 was able to achieve one further order of magnitude higher precision $\Gamma=0.876\pm{0.045}\pm{0.026} MeV$~\cite{E835chic1}.

The combination of the much better mass resolution with the ability to detect hadronic final states which have up to two orders of magnitude higher branching fractions than the $\gamma\gamma$ channel will permit high precision investigations of many open issues in the charmonium system. For instance, our knowledge of the ground state ($\eta_c$) is surprisingly poor. The existing data do not present a consistent picture, and only a small fraction of the total decay width has been measured via specific decay channels. Furthermore, radial excitations are not simple recursion of the ground state, as observed in the hadronic decays of the $\Psi$ states. In particular, the existing data on the first radial excitation of the ground state ($\eta_c '$) leave a lot to be desired. Its discovery by BELLE~\cite{BELLEeta'} was an 8$\sigma$ deviation from the initial claims of the Crystal Ball~\cite{CBeta'}. Furthermore, the existing data on the mass only have an accuracy of 4 MeV and there is a 50\% uncertainty to the width. These results are only marginally consistent with most predictions.

The singlet-P resonance ($h_c$) is of extreme importance to determine the spin dependent component of the $q \bar q$ potential. Only two decay channels have been observed for this state, and the PDG~\cite{PDG} omitts this from the summary table stating that it needs confirmation. Due to the narrow width of this state ($\Gamma<1 MeV$) only $\bar p p$ formation experiments similar to those proposed for PANDA will be able to measure the width and perform systematic investigations of the decay modes.

The energy region above the $\bar D D$ threshold has until recently been very poorly explored. Since 2003 a number of narrow states in this mass region have been observed by BELLE, BABAR and CLEO. Many basic quantities of these states have yet to be determined. Furthermore, precision measurements of all $^1D$ and $^3D$ states is required to distinguish between models that have different describtions of the nature of these states.

\section{The PANDA Detector}

FAIR will include a storage ring for beams of phase space cooled antiprotons with unprecedented quality and intensity~\cite{HESR}. The antiprotons will be collected with an average rate of about 10$^7$/s and then stochastically cooled and stored. After $5\times 10^{10}$ antiprotons have been produced, they will be transferred to the 
High Energy Storage Ring HESR where internal experiments in the beam momentum range 1.5 -- 15 GeV/c can be performed. Electron and stochastic phase space cooling will be available to allow for experiments with either high momentum resolution of about $\sim 10^{-5}$ at reduced luminosity or at high luminosity up to $2 \times 10^{32} /cm/s$ with enlarged momentum spread ($\sim 10^{-4}$).

The PANDA detector is designed as a large acceptance multi-purpose setup. The experiment will use internal targets. It is conceived to use either pellets of frozen $H_2$ or cluster jet targets for the $\bar pp$ reactions, and wire targets for the $\bar pA$ reactions. Pellet targets such as the one in operation at WASA at COSY~\cite{WASA} shoot droplets of frozen $H_2$ with radii 20 $\mu$m with typical separations of 1 mm transversely through the beam.

This detector facility must be able to handle high rates (10$^7$ annihilations/s ), with good particle identification and momentum resolution for $\gamma,~ e,~ \mu,~ \pi,~ K$, and $p$. Furthermore, the detector must have the ability to measure $D,~ K^0_S$, and $\Lambda$ which decay at displaced vertices. Finally, a large solid angle coverage is essential for partial wave analysis of resonance states.

\begin{figure}[htb]
\begin{center}
\epsfig{file=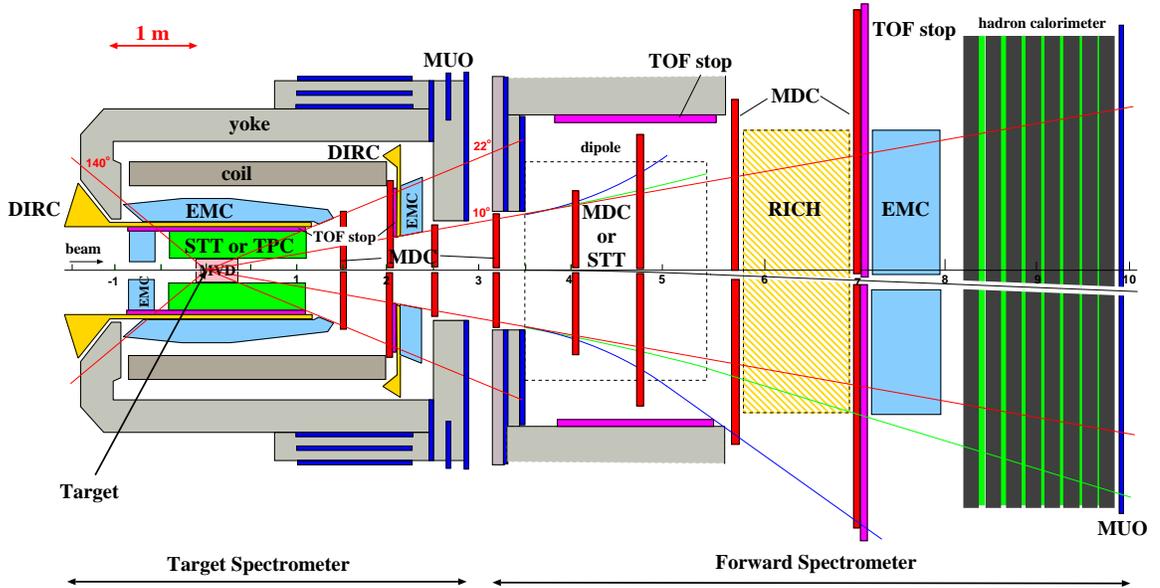, clip= , angle=0, width=\linewidth}
\caption{Schematic overview of the PANDA detector as seen from the top. The overall length is about 12 m.}
\label{fig:detector}
\end{center}
\end{figure}

In order to cope with the variety of final states and the large range of particle momenta and emission angles, associated with the different physics topics, the detector has almost 4$\pi$ detection capability both for charged particles and photons. 
 A schematic overview of the detector is given in Figure~\ref{fig:detector}.
It is divided into two sub-components, a central target spectrometer and a forward spectrometer with an overall length of 12 m of the total detector. The cylindrical structure of the target spectrometer is given by the 2 T solenoid magnetic field around the target. In sequence of increasing radii, it consists of a Micro Vertex Detector (MVD) as innermost sub-detector close to the interaction point, a central tracker built either of straw tubes (STT) or a time projection chamber (TPC) in the barrel part and mini-drift chambers (MDC) as front cap, a system of ring-imaging Cherenkov detectors for particle identification, an electromagnetic calorimeter made of $PbWO_4$ crystals, a superconducting coil, and a muon detector outside of the return yoke. The forward spectrometer consists of a 2 Tm dipole magnet with a set of multi-wire drift chambers (MDC) for tracking, a RICH detector for particle identification, electromagnetic and hadronic calorimeters for neutral and charged particles, and a muon detector as the most downstream component.

\section{Physics Reach}

The cross sections and branching ratios for many of the interesting channels to be measured are not known. Nevertheless estimates of events rates can be made in order to compare the physics reach of this experiment with other existing or planned experiments. Based upon the design luminosity of \L$=2\times10^{32}/cm^2/s$, the planned operating cycle of 6 months per year, and accounting for the duty cycle, an integrated luminosity of more than $1 fb^{-1}$ per year will be collected.  The annihilation cross section for $\bar p p \to X$ can be determined by the following formula:
\beq
\sigma_R\left( s\right) \equiv \frac{4\pi(\hbar c)^2}{(s-4m_p^2c^4)}
\frac{B_{in}B_{out}}{1+\left[ 2(\sqrt{s}-M_Rc^2)\right]^2}
\eeqn
Here $B_{in}$ and $B_{out}$ correspond to the branching ratios in and out, respectively. For instance, at the $J/\Psi$ mass the annihilation cross section is $\sigma = 1.8 \times 10^6 pb$, corresponding to about $2 \times 10^9$ $J/\Psi$ produced per year. The measured events rates must of course be scaled with the branching ratio and the detection efficiency of that channel. The corresponding rates for the $\Psi(3770)$ is of interest for the D-meson measurements, as well as other charomonium states above the $\bar DD$ threshold. In this case the branching ratio $B_{in} \equiv B(\Psi(3770)\to \bar pp)$ can be estimated by scaling with the total width of the state, \ie $B_{in}=B(J/\Psi \to \bar pp)\times \Gamma_{J/\Psi}/\Gamma_{\Psi(3770)}$. Based upon this estimate there will be $4\times 10^6$ produced $\Psi(3770)$ per year. Details of the detection efficiencies and background estimates are given in~\cite{PANDA}.

\section{Conclusion}
The PANDA collaboration has a rich and innovative program, of which only a small part could be presented here, that will be realized at the upcoming FAIR facility. The high mass resolution and integrated luminosity will open unique possibilities to do precision spectroscopy of the charmonium system.

\bigskip.

\end{document}